\documentclass[prb,twocolumn,amssymb]{revtex4}
\usepackage{graphicx}
\begin{document}

\title{Microscopic origin of local moments in a zinc-doped high-$T_{c}$
superconductor}
\author{ X. L. Qi and Z. Y. Weng}
\affiliation{Center for Advanced Study, Tsinghua University, Beijing 100084, China}
\date{{\small \today}}

\begin{abstract}
The formation of a local moment around a zinc impurity in the
high-$T_{c}$ cuprate superconductors is studied within the
framework of the bosonic resonating-valence-bond (RVB) description
of the $t-J$ model. A topological origin of the local moment has
been shown based on the phase string effect in the bosonic RVB
theory. It is found that such an $S=1/2$ moment distributes near
the zinc in a form of staggered magnetic moments at the copper
sites. The corresponding magnetic properties, including NMR spin
relaxation rate, uniform spin susceptibility, and dynamic spin
susceptibility, etc., calculated based on the theory, are
consistent with the experimental measurements. Our work suggests
that the zinc substitution in the cuprates provide an important
experimental evidence for the RVB nature of local physics in the
original (zinc free) state.

\end{abstract}

\maketitle

\section{Introduction}

The Zn substitution of the in-plane Cu$^{2+}$ ions in the high-$T_{c}$
cuprates introduces some novel properties to the system. Although a Zn$^{2+}$
ion can be considered as a nonmagnetic impurity in the CuO$_{2}$ plane,
strong magnetic signatures have been detected at the surrounding Cu sites.
The NMR and NQR experiments \cite{Julien,Alloul,Alloul2,NQR1,NQR2} have
shown the presence of staggered antiferromagnetic (AF) moments near the Zn
site, whose sum behaves like an $S=1/2$ magnetic moment as indicated by a
Curie-like $1/T$ behavior in both the spin-lattice relaxation rates and
Knight shift at low temperatures. The STM experiments have revealed \cite%
{STM} a sharp near-zero-bias peak around the Zn site, where the
quasi-particle coherent peak, appearing near the superconductivity gap in
the bulk case, is suppressed simultaneously. The zinc replacement has also
shown a strong destructive effect on $T_{c}$, and only several percentage of
zinc doping can fully destroy the bulk superconductivity \cite{Tc}.

In order to explain the STM experiment beyond a conventional nonmagnetic
impurity scattering treatment \cite{Balatsky}, the existence of a local
magnetic moment has been \emph{assumed}, whose interaction with the d-wave
nodal quasi-particles leads to a Kondo resonance peak \cite{KondoPALee,
KondoSachdev, GMZhang, Sheng}. Theoretically it remains a great challenge to
understand the microscopic origin of the magnetic moment found near a
nonmagnetic Zn impurity. In one of recent attempts, it was interpreted \cite%
{ZQWang} as due to the binding of an $S=1/2$ nodal quasi-particle to the
impurity, based on a modified mean-field theory of the $t-J$ model. The
local staggered AF moments were also explained \cite{SDW} as a local
spin-density-wave (SDW) ordering. But there still lacks a unified theory
that can self-consistently explain the free moment and local AF ordering
near the zinc impurity without suffering a magnetic instability.
Nonetheless, various approaches have at least indicated that the zinc
phenomena are quite different from those caused by nonmagnetic impurities in
a conventional superconductor and have something directly to do with the
nature of the underlying doped Mott insulators. A good understanding of the
pure strongly correlated system, therefore, is quite essential in order to
sensibly address the overall zinc impurity issue.

In this paper, we approach the zinc problem by using a microscopic
description of doped Mott insulators, in which AF correlations in
spin degrees of freedom are systematically described at various
ranges as a function of doping concentration of holes. We show
that an $S=1/2$ moment does emerge, naturally, near a zinc
impurity, which is physically originated from an RVB pair in the
original spin background. The latter becomes unpaired upon the Zn
substitution, with one of its constituent spin being removed,
together with the underlying Cu$^{2+}$ ion. In particular, we find
that such an $S=1/2$ moment cannot escape from the zinc impurity
due a topological reason. It is a consequence of the nonlocal
mutual entanglement between spin and charge degrees of freedom
known as the phase string effect. Due to this effect, a neutral
spin-$1/2$ excitation (spinon) always carries a fictitious $\pi $
fluxoid as seen by the charge carriers and, what is more, even a
vacancy on the lattice, such as a zinc impurity, also acts as a
vortex when the charge carriers are condensed. Consequently, the
condensate cannot survive in the presence of a zinc impurity
unless a spin $1/2$, which carries an antivortex, is trapped
nearby to compensate the vortex effect.

In this bosonic RVB description, we show that the induced $S=1/2$
moment around the zinc impurity distributes like staggered AF
moments due to the short-range AF correlations already present in
the spin background, which are frozen into a local AF ordering
once the direction of the moment is fixed, say, by external
magnetic fields. The calculated NMR spin-lattice relaxation rates
and uniform spin susceptibility are found to be in a systematic
agreement with the experiments. The midgap spin excitations are
also investigated, with the results consistent with the neutron experiment %
\cite{neutron}.

An important fact that we find is that all these novel properties
are already exhibited in a zinc-doped state obtained by a `sudden
approximation': simply removing a spin sitting at the zinc site
from the original pure system. A self-consistent adjustment of the
RVB background beyond such a sudden approximation only further
strengthens the local trapping of the $S=1/2$ moment as well as
local AF staggered ordering around the impurity. Therefore, a Zn
impurity in the high-$T_{c}$ cuprates constitutes a \emph{direct
}probe of the \emph{pure} system according to the present theory.
Namely, both the local $S=1/2$ moment and the staggered AF moments
`induced' by a zinc substitution truthfully mirror the nature of
the original doped Mott insulator. In this sense, the zinc
substitution simply reveals the `secrets' of local physics already
hidden in the zinc-free ground state.

The remainder of the paper is organized as follows. In Sec. II, we
present the topological reason that ensures an $S=1/2$ moment
being trapped around a zinc impurity, based on the phase string
effect in the bosonic RVB description of the $t-J$ model. Then in
Sec. III,  we present a generalized mean-field description for the
doped Mott insulator with a zinc impurity. The detailed numerical
results are given in Sec. IV. \ Finally, Sec. V is devoted to
conclusions and discussion.

\section{Topological origin of the local moment around a Zn impurity}

\subsection{Bosonic RVB framework}

We start with an effective description of the \emph{pure} system of a doped
Mott insulator, obtained \cite{PS1, PSMFd} based on the $t-J$ model in an
all-boson representation (i.e., the phase-string formalism).

The effective Hamiltonian is given by \cite{PS1, PSMFd}
$H_{eff}=H_{h}+H_{s}$, with
\begin{equation}
H_{h} =-t_{h}\sum_{\left\langle ij\right\rangle }h_{i}^{\dagger
}e^{i\left( A_{ij}^{s}-\phi _{ij}^{0}\right) }h_{j}+h.c.
\label{mfh}
\end{equation}
\begin{eqnarray}
 H_{s} &=&-\frac{J}{2}\sum_{<ij>\alpha }\Delta
_{ij}^{s}e^{i\alpha
A_{ij}^{h}}b_{i\alpha }^{\dagger }b_{j-\alpha }^{\dagger }+h.c.  \nonumber \\
&&+\frac{J}{2}\sum_{<ij>}\left| \Delta _{ij}^{s}\right| ^{2}+\lambda \left(
\sum_{i,\alpha }b_{i\alpha }^{\dagger }b_{i\alpha }-N(1-\delta )\right)
\label{mf}
\end{eqnarray}%
where $h_{i}^{\dagger }$ and $b_{i\alpha }^{\dagger }$ denote the creation
operators of bosonic holons and spinons, respectively, and $\Delta _{ij}^{s}$
is the bosonic RVB order parameter determined either self-consistently or by
minimizing the free energy. In the pure system, a uniform solution, $\Delta
_{ij}^{s}=\Delta ^{s},$ [$ij\in $ the nearest-neighbor (nn) sites], is
usually obtained \cite{PSMFd}. The Lagrangian multiplier $\lambda $ in (\ref%
{mf}) is introduced to enforce the global constraint of the total spinon
number:%
\begin{equation}
\sum_{i,\alpha }\left\langle b_{i\alpha }^{\dagger }b_{i\alpha
}\right\rangle =N(1-\delta ).  \label{sn}
\end{equation}

A unique feature in this model is the topological gauge field $A_{ij}^{s}$, $%
\phi _{ij}^{0}$, and $A_{ij}^{h}$, defined on a nn link $(ij)$, as follows:
\begin{equation}
A_{ij}^{s}=\frac{1}{2}\sum_{l\neq i,j}\left[ \theta _{i}(l)-\theta _{j}(l)%
\right] \left( \sum_{\sigma }\sigma n_{l\sigma }^{b}\right) ,  \label{eas}
\end{equation}%
\begin{equation}
\phi _{ij}^{0}=\frac{1}{2}\sum_{l\neq i,j}\left[ \theta _{i}(l)-\theta
_{j}(l)\right] ,  \label{epip}
\end{equation}%
and
\begin{equation}
A_{ij}^{h}=\frac{1}{2}\sum_{l\neq i,j}\left[ \theta _{i}(l)-\theta _{j}(l)%
\right] n_{l}^{h}.  \label{eah}
\end{equation}%
Here $n_{l\sigma }^{b}$ and $n_{l}^{h}$ are spinon and holon number
operators, respectively. By noting $\theta _{i}(l)=\mbox{Im ln $(z_i-z_l)$}$
with $z_{i}=x_{i}+iy_{i}$ representing the complex coordinate of a lattice
site $i$, one has
\begin{eqnarray}
\sum\nolimits_{C}A_{ij}^{s} &=&\pi \sum\nolimits_{l\in \Sigma _{C}}\left(
n_{l\uparrow }^{b}-n_{l\downarrow }^{b}\right),  \label{as} \\
\sum\nolimits_{C}A_{ij}^{h} &=&\pi \sum\nolimits_{l\in \Sigma _{C}}n_{l}^{h},
\label{ah}
\end{eqnarray}%
and $\phi _{ij}^{0}$ is a uniform $\pi $-flux gauge field,
satisfying
\begin{equation}
\prod_{\square }e^{i\phi _{ij}^{0}}=-1  \label{phi0}
\end{equation}
for each plaquette. It means that each holon behaves like a $\pi $
fluxoid through $A_{ij}^{h},$ which is felt by spinons in $H_{s}$
(here $C$ denotes a closed loop and $\Sigma _{C}$ is the area
encircled by it), and \emph{vice versa}. Thus holons and spinons
are mutually frustrated by each other nonlocally via the
topological fields, $A_{ij}^{s}$ and $A_{ij}^{h}$, which
represents the phase string effect hidden in the $t-J$ model.

In this description, the superconducting state is realized by the
holon condensation, $\langle h^{\dagger }\rangle \equiv h_{0}\neq
0$, while spinons remain RVB paired as characterized by $\Delta
^{s}\neq 0$. Specifically, the superconducting order parameter can
be written as \cite{PSMFd,svp}
\begin{equation}
\Delta ^{\mathrm{SC}}=\Delta ^{0}e^{i\Phi ^{s}}  \label{deltasc}
\end{equation}%
where $\Delta ^{0}\propto \Delta ^{s}h_{0}^{2}$, and $\Phi ^{s}$
is defined by
\begin{equation}
\Phi _{i}^{s}\equiv \sum_{l\neq i}\theta _{i}(l)\left( \sum_{\alpha }\alpha
n_{l\alpha }^{b}\right)  \label{phis}
\end{equation}
which describes that each spinon carries a $2\pi $ vortex in the phase of $%
\Delta ^{\mathrm{SC}}$ (known as a spinon-vortex \cite{svp}).
Since spinons form singlet RVB\ pairs, these vortices and
antivortices are generally cancelled out in (\ref{phis}) such that
the phase coherence of $\Delta ^{\mathrm{SC}}$ can be established
at low temperatures \cite{PSMFd,svp}.

A non-superconducting state, with a finite pairing amplitude $%
\Delta ^{0}$ but is short of phase coherence, can be realized at
higher temperatures where excited spinons disorder the phase $\Phi
_{i}^{s}$ according to (\ref{phis}). Such a low-temperature
pseudogap phase is called spontaneous vortex phase due to the
presence of free spinon-vortices at $T_{c}<T<T_{v}$ \cite{svp}.
The high-temperature pseudogap phase is defined at
$T_{v}<T<T_{0}$, where the holon condensation is gone such that
the pairing amplitude $\Delta ^{0}=0$, whereas the RVB order
parameter $\Delta ^{s}$ still remains finite. The latter vanishes beyond $%
T_{0}$.

The following discussion of local moments around zinc impurities
will be mainly focused in the superconducting phase and
spontaneous vortex phase, where the relation (\ref{deltasc})
generally holds with $\Delta ^{0}\neq 0$.

\subsection{Topological origin of local moments}

Now let us consider a zinc impurity added to a pure system of the doped Mott
insulator described by the bosonic RVB theory outlined above.

In the high-$T_{c}$ cuprates, chemically it is a \textrm{Cu}$^{2+}$ ion that
is replaced \cite{Chem} by a \textrm{Zn}$^{2+}$. \ A Zn-potential may be
considered as a unitary potential \cite{Balatsky} which pushes away \emph{%
both} spin and charge from the zinc site. In the framework of the $t-J$
model, a zinc impurity can be thus simply treated as an \emph{empty} site
with excluding the occupation of any electrons.

Since a zinc impurity does not change the total charge of the
system in the substitution of a \textrm{Cu}$^{2+}$ ion by a
\textrm{Zn}$^{2+}$, one may construct an effective theory by
starting with that for a \emph{pure }system and removing a neutral
spin (spinon) from the system. A heuristic procedure is to imagine
exciting a spinon at the would-be zinc site. With its spin being
`fixed', its exchange coupling with the surrounding spins is
effectively cutoff. Nor a holon can hop to this site due to the
no-double-occupancy constraint in the $t-J$ model. Whether one
removes or not such an isolated spin, the effective effect of a
zinc impurity is created at such a site.

Based on the bosonic RVB theory, corresponding to the creation of an
isolated spinon at the would-be zinc site, a $2\pi $ vortex will then appear
in the superconducting order parameter (\ref{deltasc}) via $\Phi ^{s}$ or a $%
\pi $ vortex in the holon Hamiltonian (\ref{mfh}) \ via
$A_{ij}^{s}$, in the superconducting or spontaneous vortex phase
with the condensation of holons, $\langle h^{\dagger }\rangle \neq
0.$ In other words, each zinc impurity will always induce a
nonlocal response (vortex current) from the charge condensate as
shown in Fig. 1(a).

Such a topological effect of a zinc impurity can be traced back to the
nonlocal effect in the pure system of such a doped Mott insulator, known as
the phase string effect: the motion of doped holes will always create
string-like sign defects which cannot be `repaired' at low energy in the $%
t-J$ model \cite{PS1}. Consequently spin and charge degrees of
freedom are mutually `entangled' in the pure system. An `empty'
(zinc) site then can be nonlocally perceived by the charge degrees
of freedom to result in the above vortex-like response centered at
this site.

A more rigorous derivation is to start from the phase string
representation \cite{PS1} of the $t-J$ model with the presence of
an `empty' site. Under the same RVB order parameter $\Delta
_{ij}^{s}$, one obtains essentially the same
effective Hamiltonians, (\ref{mfh}) and (\ref{mf}), except for that in (\ref%
{epip}) the summation does not include the `empty' (zinc) site,
denoted by $i_{0}$. Namely,
\begin{eqnarray*}
\phi _{ij}^{0} &\rightarrow &\phi _{ij}^{0}-\phi _{ij}^{\mathrm{Zn}} \\
\phi _{ij}^{\mathrm{Zn}} &=&\frac{1}{2}\left[ \theta _{i}(i_{0})-\theta
_{j}(i_{0})\right] .
\end{eqnarray*}%
Note that the summations in (\ref{eas}) and (\ref{eah}) should also exclude
the zinc site, which are automatically ensured since both spinons and holons
are not allowed at the site $i_{0}$, so that the definitions for $A_{ij}^{s}$
and $A_{ij}^{h}$ remain the same as in the pure system. By noting
\[
\sum\nolimits_{C}\phi _{ij}^{\mathrm{Zn}}=\pi
\]%
for a closed loop around $i_{0}$, one finds that a zinc site is bound to an
extra $\pi $-fluxoid which will always induce a vortex current in (\ref{mfh}%
) if the holons are condensed. Similarly, using the original definition of $%
\Delta ^{0}$ in (\ref{deltasc}), the effect of an empty site can
be reexpressed by the replacement $\Delta ^{0}\rightarrow \Delta
^{0}e^{i\Phi _{i_{0}}^{0}}$, with $\Phi _{i_{0}}^{0}$ giving rise
to a $2\pi $ vortex. Both are consistent with the previous
argument based on freezing a spinon at the site $i_{0}$, which
also lead to Fig. 1(a).

Now it is natural to see why a zinc impurity will generally induce a spin-$%
1/2$ around it in the superconducting and spontaneous vortex
phases. In the superconducting state, a Zn-vortex costs a
logarithmically divergent energy and thus must be `screened' by
nucleating a neutral $S=1/2$ spinon which carries an antivortex
\cite{svp} and is bound to the latter, as shown in Fig. 2b. Note
that in the pure system, an isolated spinon excitation is not
allowed in the superconducting bulk state due to the same reason,
and only spinons in bound pairs (vortex-antivortex pairs) can be
excited, which is known as the spinon confinement \cite{svp}. It
is also noted that in a different approach \cite{qhwang} based on
a similar all-boson formalism, an `unscreened' current vortex is
predicted around a zinc impurity {\it{if}} the $S=1/2$ is trapped
around which carries such a (anti)vortex. By contrast, such a
(anti)vortex is always compensated in the present approach since a
zinc impurity itself also induces a vortex as shown in Fig. 1(a).

In the spontaneous vortex phase, even though free vortices are thermally
present in the bulk in a similar fashion as in Kosterlitz-Thouless
transition, there still exists a logarithmic attraction between a Zn-vortex
and a spinon-vortex at short-range, and a bound state (although not a
confined state below $T_{c}$) between a Zn and an $S=1/2$ moment can be
still present below $T_{v}$.

Therefore, there is a fundamental topological reason for an
$S=1/2$ moment to be trapped around a zinc impurity in the bosonic
RVB theory of doped Mott insulators, in the superconducting and
spontaneous vortex phases.

\begin{figure}[tbp]
\begin{center}
\includegraphics[width=2.5in]{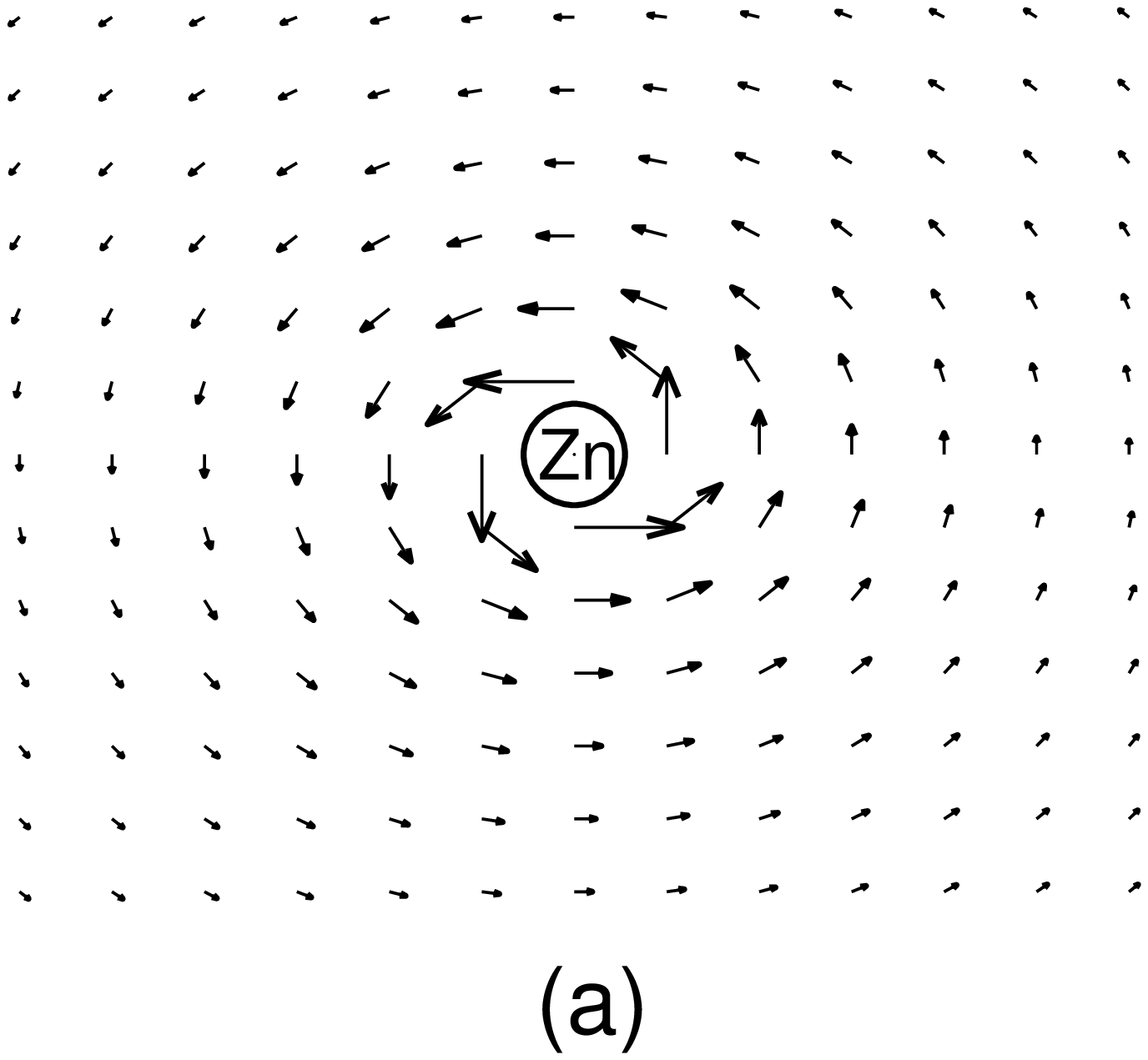}
\end{center}
\end{figure}
\begin{figure}
\begin{center}
\includegraphics[width=2.5in]{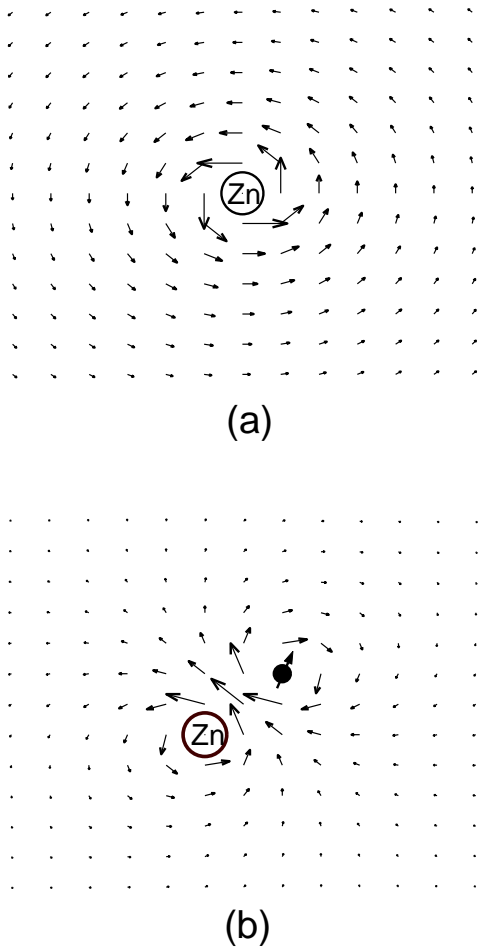}
\end{center}
\caption{(a) A vacancy (zinc impurity) always induces a
vortex-like supercurrent response in the superconducting phase due
to the phase string effect. (b) To compensate such a vortex
effect, a spinon, which carries an antivortex, has to be trapped
around the zinc impurity, giving rise to a local $S=1/2$ moment. }
\label{Topo}
\end{figure}

\section{Generalized mean-field description}

Once we have established the topological origin of the $S=1/2$ moment around
a zinc impurity, a simple effective description of the system with one zinc
impurity can be developed.

Note that after trapping a spinon nearby, the vortex induced by
the zinc impurity is compensated by the antivortex carried by the
spinon, as illustrated by Fig. 1(b). Then the system is
topologically trivial at a distance sufficiently away from the
impurity where the system remains the same as the pure system. The
change of the state mainly occurs around the impurity with a
characteristic scale comparable to the spin correlation length.

We can construct a state based on the ground state $\left| \Psi
_{0}\right\rangle $ of the pure system$,$ by removing a \emph{local} spin
sitting at the site $i_{0},$ denoted by
\begin{equation}
\left| \Psi _{0}\right\rangle _{\mathrm{Zn}}\equiv \hat{P}_{i_{0}}\left|
\Psi _{0}\right\rangle .  \label{psi0zn}
\end{equation}%
To leading order approximation, $\left| \Psi _{0}\right\rangle _{\mathrm{Zn}}
$ may be regarded as a `sudden approximation' of the true ground state $%
\left| \Psi \right\rangle _{\mathrm{Zn}}$ in the presence of a
zinc$.$ Both have the same spin and charge quantum numbers. By
`suddenly' removing a spinon at $i_{0}$, its original partner
spinon in $\left| \Psi _{0}\right\rangle $ will be left around the
site $i_{0}$ in $\left| \Psi _{0}\right\rangle _{\mathrm{Zn}}$
within the spin correlation length $\xi $. At distances larger
than $\xi $, on the other hand, $\left| \Psi _{0}\right\rangle
_{\mathrm{Zn}}$ is essentially the same as $\left| \Psi
_{0}\right\rangle .$ Due to the topological reason discussed in
last
section, the free spinon partner created in $\left| \Psi _{0}\right\rangle _{%
\mathrm{Zn}}$ should remain trapped around the impurity. Therefore one
expects $\left| \Psi _{0}\right\rangle _{\mathrm{Zn}}$ to have a good
overlap with the true ground state $\left| \Psi \right\rangle _{\mathrm{Zn}}$
and can smoothly evolve into the latter under a weak local perturbation. So
it is reasonable for one to take (\ref{psi0zn}) as a good variational form
for the zinc problem.

Since the holons are Bose-condensed in the superconducting and spontaneous
vortex phases which we are concerned, $A_{ij}^{h}$ in the spinon Hamiltonian
(\ref{mf}) can be  simplified as approximately describing a uniform flux
with a strength
\begin{equation}
\sum\nolimits_{\square }A_{ij}^{h}\simeq \pi \delta   \label{uflux}
\end{equation}%
per plaquette.

Under the condition (\ref{uflux}), the spinon Hamiltonian $H_{s}$ in (\ref%
{mf}) can be straightforwardly diagonalized as \cite{PSMFd}
\begin{equation}
H_{s}=\sum_{m,\alpha }E_{m}\gamma _{m\alpha }^{\dagger }\gamma _{m\alpha }+%
\mathrm{const.,}  \label{MFd}
\end{equation}%
by a Bogoliubov transformation%
\begin{equation}
b_{i\sigma }=\sum_{m}w_{m\sigma }(i)\left( u_{m}\gamma _{m\alpha
}-v_{m}\gamma _{m-\alpha }^{\dagger }\right)  \label{Bogolubov}
\end{equation}%
with $\left| u_{m}\right| =\sqrt{\frac{\lambda +E_{m}}{2E_{m}}}$ and $\left|
v_{m}\right| =\sqrt{\frac{\lambda -E_{m}}{2E_{m}}}$. A detailed treatment of
this mean-field state in a self-consistent way can be found in Refs. \cite%
{PSMFd,wqchen}.

Based on the above mean-field description, the zinc-free RVB ground state $%
\left| \Psi _{0}\right\rangle $ is defined by $\gamma _{m\alpha }\left| \Psi
_{0}\right\rangle =0.$ Now we construct the states with one zinc being added
to the system as discussed at the beginning of this section.

Firstly, according to the sudden approximation, the trial state with a zinc
at site $i_{0}$ may be obtained by annihilating a \emph{bare} spinon at site
$i_{0}$ with a spin index, say, $-\sigma $:%
\begin{equation}
\left| \Psi _{0}\right\rangle _{\mathrm{Zn}}=b_{i_{0}-\sigma }\left| \Psi
_{0}\right\rangle .  \label{gstrial}
\end{equation}%
Then, we allow the bosonic RVB order parameter $\Delta _{ij}^{s}$ to be
adjustable around the zinc site to further minimize the total energy. Thus
the final trial Zn state can be constructed in the following form
\begin{equation}
\left| \Psi \right\rangle _{\mathrm{Zn}}=Cb_{i_{0}-\sigma }\left| \Psi
_{0}[\Delta _{ij}^{s}]\right\rangle .  \label{gsmf}
\end{equation}%
Here $C$ is a normalization constant and $\left| \Psi _{0}[\Delta
_{ij}^{s}]\right\rangle $ is the ground state of the Hamiltonian (\ref{mf})
under a fixed form of the RVB order parameter $\Delta _{ij}^{s}$, which will
generally deviate from the uniform $\Delta ^{s}$ (of the pure system) around
the zinc site.

In order to address the dynamic and thermodynamical properties, one needs to
further determine the elementary excitations based on $\left| \Psi
\right\rangle _{\mathrm{Zn}}$ defined in (\ref{gsmf}). We take the following
steps to make the construction. Firstly, by using the Bogoliubov
transformation (\ref{Bogolubov}), one has%
\begin{eqnarray}
Cb_{i_{0}-\sigma }\left| \Psi _{0}\right\rangle &=&C\sum_{m}w_{m\sigma
}\left( i_{0}\right) v_{m}\gamma _{m\sigma }^{\dagger }\left| \Psi
_{0}\right\rangle  \nonumber \\
&\equiv &f_{0\sigma }^{\dagger }\left| \Psi _{0}\right\rangle ,  \label{f0}
\end{eqnarray}%
with $f_{0\sigma }^{\dagger }=C\sum_{m}w_{m\sigma }\left( i_{0}\right)
v_{m}\gamma _{m\sigma }^{\dagger }$ and $C\equiv \left( \sum_{m}\left|
w_{m\sigma }\left( i_{0}\right) \right| ^{2}v_{m}^{2}\right) ^{-1/2}.$
Secondly, define a new class of single spinon creation operators $f_{n\sigma
}^{\dagger }$ as a linear combination of $\gamma _{m\sigma }^{\dagger }$'s,
which satisfies
\begin{equation}
f_{n\sigma }^{\dagger }=\sum_{m}F_{nm}^{\sigma }\gamma _{m\sigma }^{\dagger
},\;\sum_{m}\left( F_{lm}^{\sigma }\right) ^{\ast }F_{nm}^{\sigma }=\delta
_{l,n}  \label{constra}
\end{equation}
with $F_{0m}^{\sigma }\equiv Cv_{m}w_{m\sigma }\left( i_{0}\right) $ such
that $f_{0\sigma }^{\dagger }$ is consistent with the definition in (\ref{f0}%
). A proper $F$ can be then obtained by re-diagonalizing the Hamiltonian (%
\ref{MFd}): $H_{s}=\sum_{n\neq 0,\alpha }\widetilde{E}_{n}f_{n\alpha
}^{\dagger }f_{n\alpha }+\mathrm{const.,}$ under a constraint $\sum_{\alpha
}f_{0\alpha }^{\dagger }f_{0\alpha }=1,$ with $\tilde{E}_{n}$ as the
`renormalized' spectrum in the presence of a zinc impurity.

Then the ground state with a zinc is simply given by $\left| \Psi
\right\rangle _{\mathrm{Zn}}=f_{0\sigma }^{\dagger }\left| \Psi
_{0}\right\rangle $ and the orthogonal (mean-field) excitation states are
constructed by the creational operators $f_{n\sigma }^{\dagger }$ ($n\neq 0$%
) as follows
\begin{equation}
\left| \left\{ \nu _{n\alpha }\right\} \right\rangle _{\mathrm{Zn}}\equiv
\prod\limits_{n(\neq 0),\alpha }\left( f_{n\alpha }^{\dagger }\right) ^{\nu
_{n\alpha }}\left| \Psi \right\rangle _{\mathrm{Zn}}.  \label{exi}
\end{equation}%
where $\nu _{n\alpha }$ denotes the occupation number at the state labelled
by ($n,\alpha $).

Physically, $f_{0\sigma }^{\dagger }$ in this approach is treated as a
projection operator, from the original ground state to the zinc-doped ground
state, while $f_{n\alpha }^{\dagger }$ ($n\neq 0$) creates spinon
excitations which are ensured to be orthogonal to $\left|\Psi\right\rangle_{%
\mathrm{Zn}}$. The presence of a zinc impurity at site $i_{0}$ will be
enforced by the constraint $\sum_{\alpha }f_{0\alpha }^{\dagger }f_{0\alpha
}=1.$

Finally, it is noted that instead of treating $\Delta _{ij}^{s}$ as
unrestricted parameters in $\left| \Psi _{0}[\Delta _{ij}^{s}]\right\rangle
, $ we shall assume a simple site-dependence for $\Delta _{ij}^{s}$ ($ij\in $
nn sites) in the following variational calculation, which is given by
\begin{equation}
\left( \Delta _{ij}^{s}\right) _{nn}=\Delta ^{s}\left[ 1-\left(
1-p_{0}\right) e^{-\frac{\left( \left| i-i_{0}\right| +\left| j-i_{0}\right|
\right) ^{2}}{4R^{2}}}\right] .  \label{dij}
\end{equation}%
Here the bulk value $\Delta ^{s}$ is decided self-consistently in the
zinc-free system \cite{PSMFd,wqchen}. The zinc at site $i_{0}$ will
influence $\left( \Delta _{ij}^{s}\right) _{nn}$ within a radius $R$ with $%
p_{0}$ determining the strength. The Lagrangian multiplier $\lambda $ in the
mean-field Hamiltonian (\ref{mf}) will be kept at the zinc-free value to
ensure that the state remains the same at distances far away from the zinc
site. The parameter $p_{0}$ will be decided by enforcing the global
constraint for the total spinon number:
\begin{equation}
\sum_{i,\alpha }\left\langle b_{i\alpha }^{\dagger }b_{i\alpha
}\right\rangle =N(1-\delta )-1  \label{sn1}
\end{equation}%
[compared to (\ref{sn})] in the presence of a zinc impurity with a given $R$%
.. In the following, most of results will be discussed for the case of $R=1$,
i.e., within the sudden approximation, and then the stability of the results
will be checked by allowing the variation of $R$.

\section{Physical Consequences}

Based on the above-constructed ground state and excited states, we can
straightforwardly calculate various related physical properties, in the
presence of a zinc impurity, by using the mean-field scheme similar to the
pure system. The results are presented below.

\begin{figure}[tbp]
\begin{center}
\emph{\includegraphics[width=2.5in]
{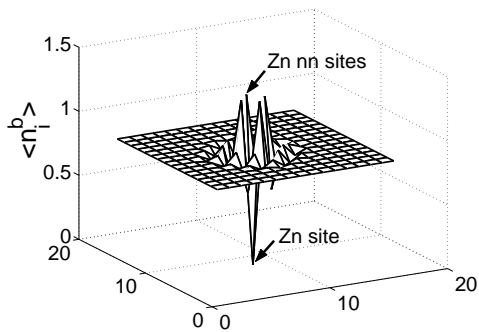} }
\end{center}
\caption{Spinon density distribution $\left\langle n_{i}^{b}\right\rangle
=\left\langle \sum_{\protect\sigma }b_{i\protect\sigma }^{\dagger }b_{i%
\protect\sigma }\right\rangle $ around the zinc impurity. The lattice is $%
16\times 16$ with doping $\protect\delta =0.125.$ }
\label{Ndis}
\end{figure}

\emph{Local }$S=1/2$ \emph{moment.---}As noted before, the ground state $%
\left| \Psi \right\rangle _{\mathrm{Zn}}$ defined in (\ref{gsmf}) differs
from the pure state $\left| \Psi _{0}\right\rangle $ by a spin $1/2$. The
change of the spinon distribution around the zinc in the ground state (\ref%
{gsmf}) is numerically determined, as illustrated in Fig. \ref{Ndis}, where
the hole concentration is fixed at $\delta =0.125,$ with the lattice size $%
16\times 16$ and $R=1$ chosen in (\ref{dij}). Compared to the zinc-free mean
value $1-\delta $ at a distance far away from the zinc site, the local
density of spinons changes within a finite length scale near the zinc site
as shown in Fig. \ref{Ndis}, which accounts for the distribution of an
unpaired spinon of $S=1/2$ in (\ref{gsmf}) created by the zinc substitution.
Note that the spinon density is enhanced in the sublattice opposite to that
of the zinc site, indicating that the unpaired spinon mainly stays there
which reflects the fact that the underlying bosonic RVB pairing in (\ref%
{gsmf}) only involves spins at different sublattices \cite{wqchen}. It
implies an AF spin configuration induced around the zinc as shown below.

\begin{figure}[tbp]
\begin{center}
\emph{\includegraphics[width=2.5in]
{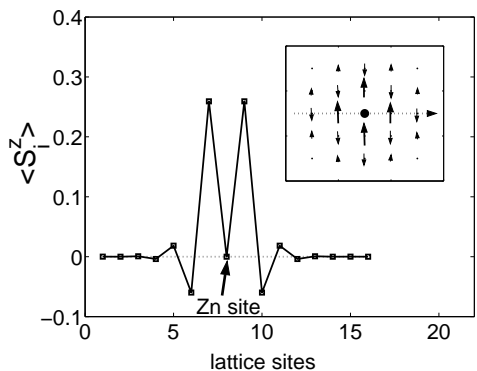} }
\end{center}
\caption{The distribution of $\left\langle S_{i}^{z}\right\rangle $ near the
zinc impurity, with the scan along the dashed direction shown in the inset,
where the zinc site is marked by the filled circle. }
\label{Sdis}
\end{figure}

\begin{figure}[tbp]
\begin{center}
\emph{\includegraphics[width=2.5in] {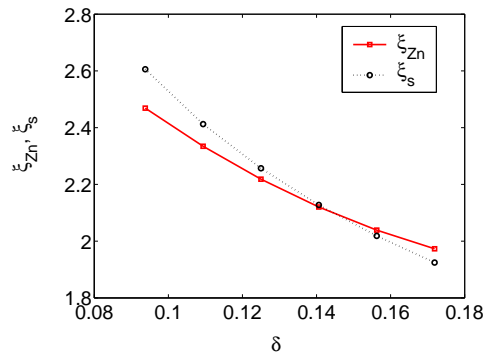}}
\end{center}
\caption{The size of the spatial distribution of the local moments, $\protect%
\xi _{Zn}$ (solid), and the bulk spin correlation length, $\protect\xi _{s}=%
\protect\sqrt{2/\protect\delta \protect\pi }$ (dotted), vs. doping $\protect%
\delta $. }
\label{CorNh}
\end{figure}

\emph{Staggered moments.---}Corresponding to the above spatial distribution
of the $S=1/2$ moment, staggered (AF) moments are further shown in Fig. \ref%
{Sdis} around the zinc site. Note that the spin rotational symmetry breaking
in Fig. \ref{Sdis} is because we fix $-\sigma =\downarrow $ in $\left| \Psi
\right\rangle _{\mathrm{Zn}}$ such that the total spin change upon a zinc
substitution is $\Delta S^{z}=1/2.$ The state $\left| \Psi \right\rangle _{%
\mathrm{Zn}}$ in general is a spin doublet according to its definition in (%
\ref{gsmf}) (if $\left| \Psi _{0}\right\rangle $ is spin singlet). The
length scale $\xi _{Zn}$ for the distribution of the local AF moments is
essentially decided by the spin-spin correlation length in the original RVB
ground state, $\xi _{s}\approx a\sqrt{2/\pi \delta }$(Ref.\cite{wqchen}).
Fig. \ref{CorNh} shows the doping dependence of the scale $\xi _{Zn}$ for
the distribution of the local spin moments, defined by fitting $\left\langle
S_{i}^{z}\right\rangle \simeq (-1)^{i}S_{0}\exp (-|i-i_{0}|^{2}/\xi
_{Zn}^{2})$. Indeed we find $\xi _{Zn}\simeq \xi _{s}.$ For a hole doping $%
\delta =0.125$, $\xi _{Zn}\simeq 2.22a$. For a typical zinc doping in
experiment, $\delta _{Zn}=0.03$, the average distance between two zincs is $%
d_{Zn}\approx a/\sqrt{\delta _{Zn}}=7.07a$, which is larger than $\xi _{Zn}$
at $\delta =0.125$. In this dilute case, the correlation effect among
different zinc impurities can be neglected and one may only focus on the
single zinc effect.

\begin{figure}[tbp]
\begin{center}
\emph{\includegraphics[width=2.5in]
{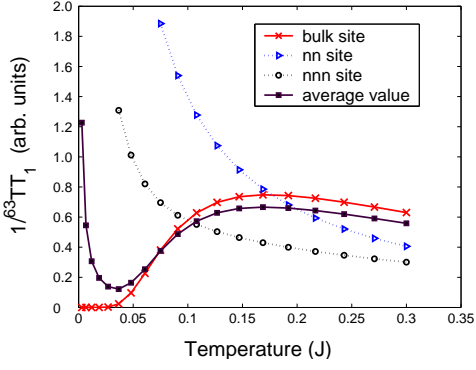}}
\end{center}
\caption{Contributions to $1/^{63}T_{1}T$ from different sites are
calculated. Solid curve with crosses: from the site far from the zinc
impurity; Dashed curve with triangles: the nn site near the zinc; Dashed
curve with circles: the next nearest neighbor (nnn) site near the zinc;
Solid curve with squares: average over all sites in a $16\times 16$ lattice
with one zinc (see text).}
\label{1ovTT1}
\end{figure}

\begin{figure}[tbp]
\begin{center}
\emph{\includegraphics[width=2.5in]
{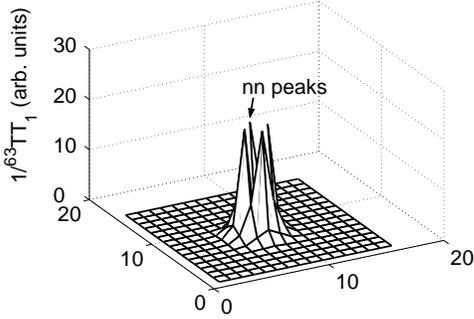}}
\end{center}
\caption{Distribution of the contributions to $1/^{63}T_{1}T$ from
individual sites near the zinc impurity, at temperature $T=0.0067J$.}
\label{TT1dist}
\end{figure}

\emph{Low-energy spin excitations.---}One can further examine the dynamic
properties of the induced local AF moments. One physical quantity is the
spin-lattice relaxation rate of $^{63}\mathrm{Cu}$ measured in NMR/NQR
experiments, which is decided by the imaginary part of the spin correlation
function $\chi \left( q,\omega \right) $ as follows
\begin{equation}
\frac{1}{^{63}T_{1}T}=\left. \sum_{q}\frac{A^{2}\left( q\right) \mathop{\rm
Im}\chi \left( q,\omega _{N}\right) }{\omega _{N}}\right| _{\omega
_{N}\rightarrow 0}  \label{NMR}
\end{equation}%
Here the structure factor $A\left( q\right) =A+2B\left( \cos q_{x}a+\cos
q_{y}a\right) $ and $A=-4B$ when the applied magnetic field is orthogonal to
the $\mathrm{CuO}_{2}$ plane \cite{Julien}. So the main contribution will
come from the AF correlations near the AF\ momentum $(\pi ,\pi ).$ Fig. \ref%
{1ovTT1} shows the theoretical calculations. In the pure system of the
bosonic RVB state, a pseudogap opens up in the spin excitations, resulting a
suppression of $1/^{63}T_{1}T$ at low temperature (solid line with crosses
in Fig. \ref{1ovTT1}). However, for the zinc-doped system, the `spin gap'
shown in Fig. \ref{1ovTT1} is filled up by a Curie-type contribution, $%
1/^{63}T_{1}T\varpropto 1/T$, due to the presence of a free moment with a
staggered distribution around the zinc site. The spatial distribution of $%
1/^{63}T_{1}T$ is given in Fig. \ref{TT1dist} at a low temperature ($%
T=0.0067J)$, which clearly shows that the Curie-type signals are located
near the zinc site, with its maximum at the nn sites of the zinc, consistent
with the distributions of the staggered moments in Fig. \ref{Sdis}.

It is noted that the calculation is performed at a $16\times 16$ lattice
with only one zinc, with an effective zinc doping concentration equal to $%
1/256\sim 0.004$. In the realistic case with more zinc impurities, the
intensity of $1/^{63}T_{1}T$ on average is expected to be proportional to
the density of zinc impurities at low temperatures, so long as the zinc
concentration is not too large such that the single-zinc approximation used
above is still valid. Thus one should multiply a factor $5\sim 8$ to the
low-temperature part of $1/^{63}T_{1}T$ for the averaged case in Fig. \ref%
{1ovTT1} (solid curve with squares) in order to compare with the
experimental case with $\delta _{\mathrm{Zn}}=0.02\sim 0.03.$

\begin{figure}[tbp]
\begin{center}
\emph{\includegraphics[width=2.5in]
{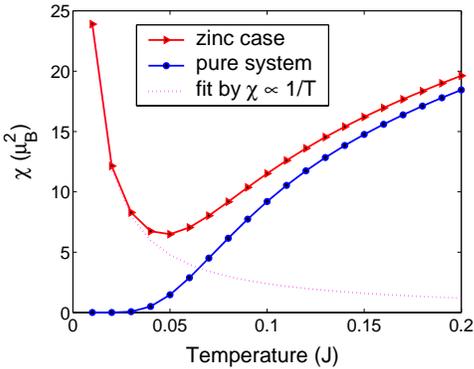}}
\end{center}
\caption{Uniform spin susceptibility in the pure system is shown by the
solid curve with full circles; The case with one zinc is illustrated by the
solid curve with triangles; The dashed curve is a fit by$\ \protect\chi %
=0.2390/T$.}
\label{Unif}
\end{figure}

Furthermore, the uniform spin susceptibility can be directly calculated by $%
\chi _{u}\propto \mathop{\rm Re}\chi \left( q=0,\omega \rightarrow 0\right) $%
.. $\chi _{u}$ also show a pseudogap behavior at low temperatures in the pure
state, which is replaced by the Curie-like $1/T$ behavior near the zinc site
due to the contribution from the local moment as shown in Fig. \ref{Unif}.

\begin{figure}[tbp]
\begin{center}
\emph{\includegraphics[width=2.5in]
{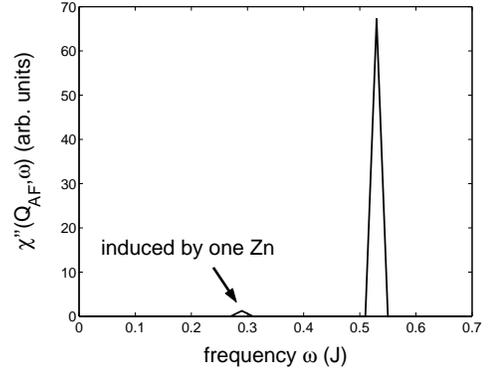}}
\end{center}
\caption{Dynamic spin susceptibility at AF wavevector $Q_{\mathrm{AF}%
}=\left( \protect\pi ,\protect\pi \right) $ with $\protect\delta =0.125$.
The high-energy resonancelike peak at $E_{g}\simeq 0.53J$ is from the bulk
(zinc free) state, while the low-energy excitations indicated by the arrow
are new ones induced by the zinc impurity. }
\label{dyn}
\end{figure}

\emph{Dynamic spin susceptibility. ---}The imaginary part of the dynamic
spin susceptibility, $\mathop{\rm Im}\chi \left( q,\omega \right) ,$ can be
directly measured by inelastic neutron scattering. In the bosonic RVB mean
field theory, a resonance peak at the AF\ wave vector $Q_{\mathrm{AF}%
}=\left( \pi ,\pi \right) $ is present in the dynamic spin correlation
function in the superconducting phase. For the hole doping $\delta =0.125$,
we obtain the resonance energy $E_{g}\approx 0.53J$ \cite{wqchen}. Upon the
zinc doping, the energy $E_{g}$ of the resonance peak has changed little
(see Fig. \ref{dyn}) in the bulk. But a zinc impurity does induce some new
states at lower energies as shown in Fig. \ref{dyn}, which reflects the
modified spin excitation spectrum near the zinc, accompanying the emergence
of a local moment. Note that the weight of such zinc-induced modes in Fig. %
\ref{dyn} should be enhanced with a finite concentration of the zincs.

\emph{Effect from the holon redistribution.---}So far the mean-field results
are obtained based on the assumption that the Bose condensed holons are
uniformly distributed in space. But in the presence of a zinc, the holon
density near the zinc impurity should be generally suppressed, to be
consistent with that the spinon density increases around the zinc site due
to the no double occupancy constraint. In the bosonic RVB theory, since the
holons will influence the spinon part by the gauge field $A_{ij}^{h}$
defined in (\ref{ah}), the suppression of the holon density around a zinc
will cause additional effects on the local spin dynamics, which is
considered below.

\begin{figure}[tbp]
\begin{center}
\emph{\includegraphics[width=2.5in] {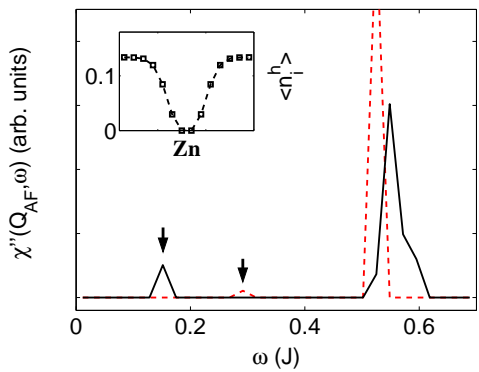}}
\end{center}
\caption{Dynamic spin susceptibility at $Q_{\mathrm{AF}}=(\protect\pi ,%
\protect\pi )$ is shown as the solid curve when the holon density is
suppressed locally around the zinc (see the inset). For comparison, the
dashed curve illustrates the case for a uniform holon density distribution.}
\label{DynNh}
\end{figure}

\begin{figure}[tbp]
\begin{center}
\emph{\includegraphics[width=2.5in] {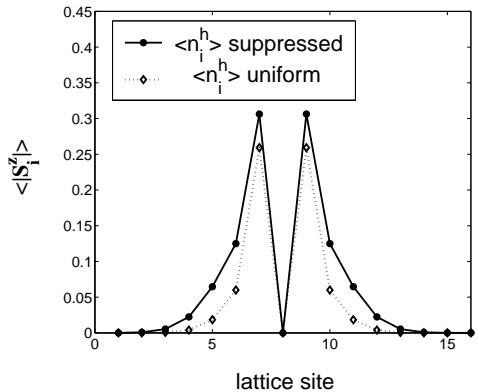}}
\end{center}
\caption{The distribution of $|\langle S_{i}^{z}\rangle |$, corresponding to
the profile of the holon density shown in the inset of Fig. \ref{DynNh}, is
plotted as the solid curve, as compared to the dashed curve for the uniform
holon distribution. }
\label{snh}
\end{figure}

As shown in Fig. \ref{DynNh}, the intensity of the zinc-induced low-energy
spin excitations seen in Fig. \ref{dyn} will be enhanced with its energy
scale further reduced if a holon density reduction is taken into account
with a profile given by the inset. Meantime, the local staggered moments
will also be increased (Fig. \ref{snh}) under the same holon distribution.
Although the above calculations are not based on a self-consistent scheme,
which is generally quite difficult, all the important features found
previously in the uniform profile of the holon distribution are kept
unchanged qualitatively, except for that the anomalies are further
strengthened due to the reduction of the holons which makes the area near
the zinc closer to the half-filling.

\begin{figure}[tbp]
\begin{center}
\emph{\includegraphics[width=2.5in]
{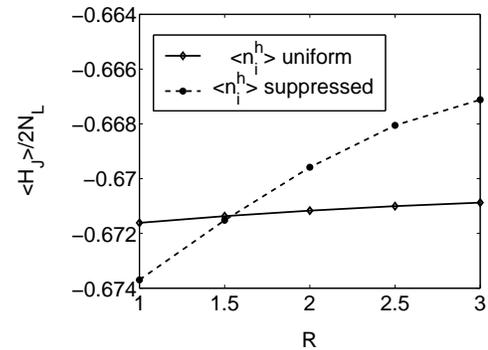}}
\end{center}
\caption{The average superexchange energy $\langle H_{J}\rangle /(2N_{L})$
per bond ($N_L$ denotes the total number of lattice sites excluding the zinc
sites) as a function of the parameter $R$ in (\ref{dij}). Solid curve:
uniform holon distribution. Dashed curve: locally suppressed holon density
around the zinc indicated by the inset of Fig. 10. }
\label{Stab}
\end{figure}

\emph{Local stability of the mean-field theory.---}The stability of the
results discussed above based on the conjectured ground state (\ref{gsmf})
can be further examined by tuning the variational order parameter defined in
(\ref{dij}). By changing the parameter $R$, the range of the zinc effect on
the RVB parameter $\left( \Delta _{ij}^{s}\right) _{\mathrm{nn}}$ in (\ref%
{dij}) can be continuously adjusted. As shown in Fig. \ref{Stab}, the
superexchange energy $\langle \tilde{H}_{J}\rangle _{\mathrm{Zn}}$ is found
to monotonically grow with the increase of $R$ for both uniform and
non-uniform profiles of the holon distribution, indicating that the sudden
approximation state $|\Psi _{0}\rangle _{\mathrm{Zn}}\simeq |\Psi
(R=1)\rangle _{\mathrm{Zn}}$ remains locally stable, which prevents the
local $S=1/2$ moment from leaking far away from the zinc impurity. When the
holon redistribution discussed above is considered (see the dashed curve in
Fig. \ref{Stab}), this local stability is further strengthened, since the
frustration effect on the spin dynamics, which comes from the holon motion,
is weakened around the zinc impurity due to the reduction of the holon
density.

\section{conclusion}

In this paper, we have developed a microscopic description of the
zinc doping effect in the cuprate superconductors based on an
effective theory, \emph{i.e.,} the bosonic RVB theory, of the
$t-J$ model. The unusual effects of the zinc doping observed in
the experiments have been explained as a direct consequence of the
mutual nonlocal entanglement between the spin and charge degrees
of freedom (\emph{i.e.,} the phase string effect) in the doped
Mott insulator. Once a local moment is formed due to such a
topological mechanism, the effect of zinc doping can be simply
described by a `sudden approximation', which directly translates
the short-range RVB pairing, present in the spin background state
of the pure system, to the AF spin distribution around a zinc
impurity, once the latter is introduced to the system. The NMR
spin relaxation rates, uniform spin susceptibility, and the
induced low-lying spin excitations, obtained within the sudden
approximation and beyond, have consistently painted a unified
picture of spin correlations near the zinc, which is shown to be
stable locally and in an overall agreement with the experimental
observations. In particular, the theory predicts that the range of
the distribution for the local moment is inversely proportional to
the square root of doping concentration.

In this work, an important property in the zinc problem has not
been discussed so far. That is the behavior of quasi-particle
excitations and the single-electron tunnelling properties. In the
bosonic RVB theory, an electron is composed of a holon and a
spinon, together with a nonlocal phase string
factor \cite{PSQP}. In the superconducting phase, it has been shown \cite%
{PSQP} that a quasi-particle is stable due to a confinement of these
holon-spinon and phase-string objects. So the overall low-energy
single-particle spectral function is expected to be the same as in a d-wave
BCS theory. However, the composite structure is predicted to be seen at
high-energies, which may explain the `coherent peak' in the antinodal regime
in the pure system as due to the spinon excitation \cite{PSQP}. In the zinc
doping case, due to the trapping of a free moment (spinon) near the zinc,
the high-energy spinon contribution in the single-particle spectral function
can become a zero-biased mode. Detailed study of this property needs the
knowledge about the charge degree of freedom as well as the confinement in
the superconducting phase, which is beyond the scope of this work. This
issue and other problems, like how the superconducting transition
temperature will be affected by the local moments, will be further
investigated within the bosonic RVB description in future.

\begin{acknowledgments}
We acknowledge helpful discussions with W. Q. Chen, Z. C. Gu, T. Li, H.
Zhai, F. Yang and Y. Zhou. The work is supported by the NSFC grants, and the
grant no. 104008 and SRFDP from MOE of China.
\end{acknowledgments}

\end{document}